\documentclass[showpacs,twocolumn,superscriptaddress]{revtex4}
\usepackage{epsfig}
\renewcommand{\epsilon}{\varepsilon}
\begin{document}

\def\s{\sigma}
\def\t{\tau}
\def\d{\delta}
\def\be{\begin{equation}}
\def\ee{\end{equation}}
\def\o{\omega}
\def\bea{\begin{eqnarray}}
\def\eea{\end{eqnarray}}
\renewcommand\L{\mathcal{L}}
\def\G{\mathcal{G}}
\newcommand\egal{&\!\!=\!\!&}

\def\arrowsite{\begin{picture}(45,5)(-2,-2)
\put(0,0){\circle*{5}}
\put(2.5,0){\vector(1,0){10}}
\put(12.5,0){\line(1,0){15}}
\put(37.5,0){\vector(-1,0){10}}
\put(40,0){\circle*{5}}
\end{picture}}

\def\arrowroot{\begin{picture}(45,5)(-2,-2)
\put(0,0){\circle*{5}}
\put(2.5,0){\vector(1,0){10}}
\put(15,0){\circle{5}}
\put(25,0){\circle{5}}
\put(37.5,0){\vector(-1,0){10}}
\put(40,0){\circle*{5}}
\end{picture}}

\title{Logarithmic Conformal Field Theory and Boundary Effects in the Dimer Model
}

\date{\today}

\author{N. Sh. Izmailian}
\affiliation{Institute of Physics, Academia Sinica, Nankang,
Taipei 11529, Taiwan}
\affiliation{Yerevan Physics Institute,
Alikhanian Brothers 2, 375036 Yerevan, Armenia}
\affiliation{National Center of Theoretical Sciences at Taipei,
Physics Division, National Taiwan University, Taipei 10617,
Taiwan}
\author{V. B. Priezzhev}
\affiliation{Bogolubov Laboratory of
Theoretical Physics, Joint Institute for Nuclear Research, 141980
Dubna, Russia}
\author{Philippe Ruelle}
\affiliation{Institut de Physique Th\'eorique,
Universit\'e catholique de Louvain, B-1348
Louvain-La-Neuve, Belgium}
\author{Chin-Kun Hu$^*$}
\affiliation{Institute of Physics, Academia Sinica, Nankang,
Taipei 11529, Taiwan}

\begin{abstract}
We study the finite-size corrections of the dimer model on
$\infty \times N$ square lattice with two different boundary
conditions: free and periodic. We find that the finite-size
corrections in a crucial way depend on the parity of $N$; we also
show that such unusual finite-size behavior can be fully explained
in the framework of the $c=-2$ logarithmic conformal field theory.
\end{abstract}
\pacs{05.50+q, 75.10-b}

\maketitle

Universality, scaling, and exact finite-size corrections in
critical systems have attracted much attention in recent decades
\cite{ufssf,ih01,Ivasho,ioh03}. It has been found that critical
systems can be classified into different universality classes so
that the systems in the same class have the same set of critical
exponents, universal finite-size scaling functions, and amplitude
ratios \cite{ufssf,ih01}. Two-dimensional critical systems are
parameterized by the central charge $c$ \cite{Belavin}, directly
related to the finite-size corrections to the critical free energy
\cite{Blote}.

In this Letter, we address this question for dimers defined on a
square lattice, with three main purposes: (i) we dissipate the
confusion existing in the literature about the value of the
central charge \cite{TzengWu,Cha} due to the (mis)interpretation
of the finite-size corrections in terms of the central charge
rather than the effective central charge; (ii) we give a bijection
of dimer coverings with spanning tree and Abelian sandpile model,
which not only allows a proper understanding of the dimer model
but proves also very useful to calculate finer effects, like the
change of boundary conditions; (iii) using this bijection, we
clarify and explain why a change of parity of the lattice size
causes  a change of the effective central charge but not of the
central charge itself.

The dimer problem on planar lattices belongs to the class of
``free-fermion'' models \cite{FanWu}. Its solution has been obtained
with the Pfaffian approach \cite{Kastelyn} and then reproduced by
a variety of methods \cite{Baxter}. In contrast to the statistics
of simple particles, the critical behavior of the dimer model is
strongly influenced by the structure of the lattice space. The
square lattice dimer model is critical with algebraic decay of
correlators \cite{Fisher2}. For the dimer model on the anisotropic
honeycomb lattice, which is equivalent to five-vertex model on the
square lattice \cite{Wu5}, the free energy exhibit a KDP-type
singularity. For the triangular and some decorated lattices, the
dimer model exhibit Ising-type transitions \cite{Wu6}. Thus, it
appears that the dimer model itself has not a single critical
behavior, but several critical behaviors associated with different
universality classes.

In what follows, we consider the finite-size effects for
close-packed dimers on finite square lattices, with free boundary
conditions on all sides (strip geometry), and with periodic
boundary condition in one direction (cylinder). In all cases we
find them to be consistent with a central charge $c=-2$. This
conclusion relies on a careful distinction between the central
charge $c$ and the so-called effective central charge $c_{\rm eff}
= c - 24h_{\rm min}$ \cite{itz}, which is a boundary dependent
quantity. The value $c=-2$ is further confirmed by calculating the
effect of a change of boundary conditions. We find that, in the
scaling limit, it corresponds to the insertion of a boundary
primary field of weight $-1/8$, belonging to a logarithmic
conformal field theory with $c=-2$.


{\it Finite-size analysis.} - Let us consider the dimer model on an
$M \times N$ square lattice $\L$
with $M$ rows and $N$ columns. The topology of
$\L$ is fixed by the boundary conditions: it forms a rectangle if
free boundary conditions are imposed in two directions, a cylinder
or a torus if periodic boundary conditions are chosen in only horizontal
direction, or two directions.

The partition function of the dimer model is given by
\begin{equation}
Z_{M, N}(z_v,z_h)=\sum z_v^{n_v}z_h^{n_h},
\label{PartitionFunctionDimer}
\end{equation}
where the summation is over all dimer covering configurations, $z_v$ and $z_h$
are the dimer weights in the vertical and horizontal directions, and
$n_v$ and $n_h$ are the numbers of vertical and horizontal dimers.

The partition functions of the dimer model
with the boundary conditions discussed above can all be expressed in
terms of $Z_{\alpha,\beta}(z,M,N)$ for $\alpha,\beta = 0,
\frac{1}{2}$ \cite{ioh03} with
\be
Z^2_{\alpha,\beta}(z,M,N) = \prod_{n=0}^{N-1}\prod_{m=0}^{M-1}4\Big[\textstyle{
z^2 \sin^2{\frac{\pi (n+\alpha)}{N}} +
\sin^2{\frac{\pi (m+\beta)}{M}}}\Big].
\label{zab}
\ee
Here $z = z_h/z_v$, which we set equal to 1 from now on.

The general theory about the asymptotic expansion of
$Z_{\alpha,\beta}$ for large $M,~N$ has been presented in
\cite{Ivasho}. For what follows, the asymptotic expansion of the
free energy per unit length associated to $Z_{\alpha,\beta}$ is
all we need. The result reads \cite{Ivasho} \bea
F_{\alpha,\beta}(N) \egal -\lim_{M \to \infty} {1 \over M} \ln
Z_{\alpha,\beta}(1, M, N) \nonumber\\
&& \hskip -1cm =  -\frac{2G}{\pi} N + \sum_{p=0}^\infty
\left(\frac{\pi}{N}\right)^{2p+1}\frac{2 z_{2p}}{(2p)!}
\frac{B_{2p+2}(\alpha)}{2p+2}, \label{AsymptoticExpansion1} \eea
where $z_0 = 1,  z_2 = -2/3, ...$ ; $G=0.915965$ is the Catalan
constant, and the $B_{p}(\alpha)$ are the Bernoulli polynomials,
$B_2(\alpha) = \alpha^2 - \alpha + {1 \over 6}$. These formulae
allow to compute the asymptotic expansion of the free energy $F_N$
per unit length of the $\infty \times N$ lattice for large $N$ and
for free and periodic boundary conditions.

Let us first consider the case of an infinitely long strip of width $N$ with
free boundary conditions. For $N$ even, the results of \cite{ioh03} show that
\be
F_{N,even}^{free} = {\textstyle {1 \over 2}} F_{{1 \over 2},0}(N+1)
+ \log{(1+\sqrt{2})}.
\ee
The expansion (\ref{AsymptoticExpansion1}) then gives
\be
F_{N,even}^{free} = -\frac{G}{\pi} (N+1) + \log{(1+\sqrt{2})} -
\frac{\pi}{24}\frac{1}{N} + \ldots
\label{feven}
\ee
Similar calculations for $N$ odd lead to a formula in terms of
$Z_{0,{1 \over 2}}(N+1)$, and yield
\be
F_{N,odd}^{free} = -\frac{G}{\pi} (N+1) + \log{(1+\sqrt{2})} +
\frac{\pi}{12}\frac{1}{N} + \ldots
\label{fodd}
\ee
The analogous results for the periodic case, {\it i.e.} an infinite cylinder
of perimeter $N$, read
\bea
F_{N,even}^{per} \egal -\frac{G}{\pi} N - \frac{\pi}{6}\frac{1}{N} + \ldots
\label{peven} \\
F_{N,odd}^{per} \egal -\frac{G}{\pi} N + \frac{\pi}{12}\frac{1}{N} + \ldots
\label{podd}
\eea

The free energy per unit length of an infinitely long strip of
finite width $N$ at criticality has the finite-size scaling form
\cite{Blote}
\begin{equation}
F=f_{\rm bulk} N + f_{\rm surf} + \frac{A}{N} + ...,
\label{freeenergy}
\end{equation}
where $f_{\rm bulk}$ and $f_{\rm surf}$ are, respectively, the bulk
and the surface (boundary) free energy densities, and $A$ is a
constant. Though the free energy densities $f_{\rm bulk}$ and
$f_{\rm surf}$ are not universal, the constant $A$ is universal.
The value of $A$ is related to the conformal anomaly $c$ of the
underlying conformal theory, and depends on the boundary
conditions in the transversal direction. These two dependencies
combine into a function of the effective central charge $c_{\rm
eff} = c - 24h_{\rm min}$,
\begin{eqnarray}
A \egal -{\pi \over 24}\, c_{\rm eff} = \pi \left(h_{\rm min}-\frac{c}{24}\right)
\;\; \mbox{on a strip}, \label{Astrip} \\
A \egal -{\pi \over 6}\, c_{\rm eff} = 4\pi \left(h_{\rm min}-\frac{c}{24}\right)
\;\; \mbox{on a cylinder}.
\label{Acyl}
\end{eqnarray}
The number $h_{\rm min}$ is the (chiral) conformal weight of the operator with the
smallest scaling dimension present in the spectrum of the Hamiltonian with the given
boundary conditions (for the cylinder, we assumed that this operator is scalar,
$h_{\rm min} = \bar h_{\rm min}$).

In a unitary theory, one has $h_{\rm min}=0$ on a cylinder (with periodic condition)
and on a strip with identical left and right boundary conditions, and $h_{\rm min} >
0$ otherwise. In a non-unitary theory, like the conformal theory discussed here, there
is no restriction on $h_{\rm min}$.

The finite-size corrections computed above have all the form (\ref{freeenergy}),
with the effective central charge depending on the parity of $N$, see (\ref{Astrip})
and (\ref{Acyl}). We will show that indeed the effective central charge, and not
the central charge itself, depends on the parity of $N$, because value of $h_{\rm min}$
does, due to the fact that changing the parity of $N$ in effect changes the boundary
condition.

To understand this peculiarity of the dimer model, we consider,
first on the strip then on a cylinder, the mapping of the dimer
model to the spanning tree model \cite{Temperley} and,
equivalently, the Abelian sandpile model \cite{Sand}.


{\it Dimers on a strip.} - Let us consider first the dimer model
on the rectangular lattice $\L$ of size $M \times N$ with free
boundary conditions. Since we are interested in the limit $M \to
\infty$, the parity of $M$ will not matter here. For simplicity,
we take $M$ odd, and discuss successively the cases $N$ odd and
$N$ even.

For $M$ and $N$ both odd, the bijection between dimer coverings on $\L$ with one
corner removed and spanning trees on the odd sublattice $\G \subset \L$ is well-known
\cite{Temperley,TzengWu}.

A dimer containing a site of $\G$, in blue in Figure 1, can be
represented as an arrow directed along the dimer from this site to
the nearest neighbour site of $\G$. It is easy to prove that the
resulting set of arrows generates a uniquely defined spanning
tree, rooted at the corner which had been removed from $\L$ (see
Figure 1). Since the dimers which do not contain a site of $\G$
are completely fixed by the others, one has a one-to-one
correspondence between dimer coverings on $\L$ minus a corner and
spanning trees on $\G$. This allows one to express the number of
dimer configurations by the Kirchhoff theorem as $Z =\det \Delta$
where $\Delta$ is the Laplacian matrix for spanning tree on $\G$.
As shown in \cite{Sand}, spanning trees on $\G$, rooted at a
corner, are in bijection with the configurations of the Abelian
sandpile model (ASM) on $\G$, with closed boundary conditions on
the four boundaries, the only sink (dissipative) site being the
root of the trees.

\begin{figure}
\epsfig{file=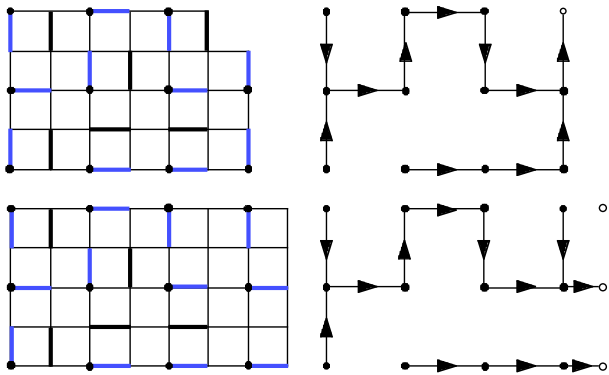,width = 8cm} \caption{(Color) Mapping
of a dimer covering to a spanning tree on the odd sublattice, for
an $M \times N=5 \times 7$ lattice (top), and a $5 \times 8$
lattice (bottom). In both cases, the solid dots represent the
sites of the odd sublattice $\G$, and the open dots are the roots
of the trees.}
\end{figure}

When $M \to \infty$, the lattice $\G$ becomes an infinitely long strip of width $N$
and, in the ASM language, closed boundary conditions on the two vertical sides.
Many independent arguments and explicit calculations all converge to a central
charge $c=-2$ for the ASM on a square lattice \cite{Sand,ruelle,PiRu}.
On the other hand, the spectrum of the ASM Hamiltonian on a slice of the strip
with closed boundary conditions at the two ends has been computed in
\cite{ruelle}. It is given by a single representation $\cal R$, reducible but
indecomposable, of the chiral algebra of a rational logarithmic conformal field
theory with $c=-2$ \cite{Gaberdiel}. The representation has two ground-states, the
identity operator and its logarithmic partner, both of conformal weight 0, so that
$h_{\rm min}=0$. The effective central charge in this sector is therefore $c_{\rm eff} = -2$,
and the general formula (\ref{Astrip}) reproduces the finite-size corrections
(\ref{fodd}).

When $M$ is odd and $N$ is even, dimer coverings exist on $\L$ without the need to
remove a corner. In this case, the above construction leads to a set of
spanning trees on the odd sublattice $\G$, where certain arrows may point out of
the lattice from the right vertical side (Figure 1). Viewing
this vertical boundary of $\G$ as roots for the spanning trees, we see that
dimer coverings on $\L$ map onto spanning trees on $\G$ which can grow from
any site of the vertical side. In turn, the spanning trees map onto the ASM
configurations with one vertical open, dissipative boundary, and the three other
closed.

In the limit $M \to \infty$, the lattice becomes an infinite strip with open and
closed boundary conditions on the two sides. In this case, the results of \cite{ruelle}
show that the ground-state of the Hamiltonian with such boundary conditions is a
primary field of conformal weight $h_{\rm min}=-1/8$. With $c=-2$, this yields $c_{\rm
eff} = 1$ and again the formula (\ref{Astrip}) correctly gives the result
(\ref{feven}).

Let us note that the bijection between the dimer coverings and the spanning
trees holds if we use the even sublattice instead of the odd one. The boundary
conditions however change. If $N$ is odd, the vertical sides (and the
horizontal ones as well for $M$ odd) become open rather than closed. The spectrum
of the corresponding Hamiltonian change, with a non-degenerate ground-state being the
identity operator \cite{ruelle}. Thus the value $h_{\rm min}=0$ remains. If $N$ is
even, the left and right boundaries, previously closed and open respectively, become
open and closed, so that the Hamiltonian remains the same, $h_{\rm min}=-1/8$.

The equivalence of the odd and even sublattices is a duality property. The two
sublattices are dual to each other, and the spanning trees on $\G_{\rm even}$ are dual
to those on $\G_{\rm odd}$. It is not difficult to check that open and closed
boundary conditions are exchanged under duality.

Thus the leading finite-size corrections for an infinitely long strip of width $N$
agree with the prediction of a $c=-2$ conformal field theory, provided one
realizes that changing the parity of $N$ genuinely changes the boundary conditions,
an effect due to the strong non locality of the dimer model. The change of
boundary conditions is not apparent in the dimer model itself, but is manifest when
one maps it onto the spanning tree model or the sandpile model.

The primary field with $h=-1/8$ can be further tested
\cite{ruelle}. If, on a rectangular lattice $M \times N$, we
remove $n$ sites from the lower boundary, the height will be $M$
or $M-1$ and will therefore take on the two parities. This has the
effect of changing the boundary condition along the lower
boundary, from closed to open, and back to closed (or vice-versa).
We have checked that the universal part of the ratio of partition
functions, after and before the removal, and expected to be equal
to $\langle \phi_h(0) \phi_h(n) \rangle \sim n^{-2h}$ where
$\phi_{h}$ is the boundary field that changes a boundary condition
from open to closed, is indeed asymptotically equal to $n^{1/4}$,
in the limit $M,N \to \infty$, supporting $h=-1/8$.


{\it Dimers on a cylinder.} - We consider here an $M \times N$
rectangular lattice $\L$ with periodic boundary condition in the
horizontal direction, so that $\L$ is a cylinder of perimeter $N$
and height $M$. As before, we will eventually take $M$ to
infinity, which makes its parity irrelevant. We choose $M$ even.
According to the discussion of the previous section, the top
boundary of the cylinder then is subjected to open boundary
conditions (in ASM terms) while the bottom one is closed. We
separate the cases $N$ odd and $N$ even. If $N$ is odd, we select
the sublattice $\G$ consisting of those sites of $\L$ having
odd-odd coordinates. It is easy to see that two columns of $\G$
will contain sites which are neighbors in $\G$ and in $\L$
(connected by horizontal bonds). Therefore a dimer may contain
zero, one or two sites of $\G$. The dimers containing no site of
$\G$ are completely fixed by the others and play no role. For the
others, we do the same construction as before. If a dimer touches
one site of $\G$, we draw an arrow directed along the dimer from
that site to the nearest neighboring site of $\G$. However, for a
dimer containing two sites of $\G$, the two arrows would point
from either site to the other, ruining the spanning tree picture.
It can nevertheless be restored in the following way.

Instead of seeing the two arrows as pointing from one site to its
neighbor, we say that they point towards roots inserted between
the neighbor sites, thus replacing the arrows \mbox{\arrowsite} by
\mbox{\arrowroot}. This in effect amounts to cut the cylinder
along the bonds of $\L$ which connect sites of $\G$, unwrapping it
into a strip, and to add columns of roots on the left and on the
right side of the strip. The new arrow configurations define
spanning trees, rooted anywhere on the left and right boundaries.
So dimer coverings on the original cylinder are mapped to spanning
trees on a strip, with open top and closed bottom horizontal
boundaries (we chose $M$ even), and open vertical boundaries.

When $M$ goes to infinity, the lattice becomes an infinite strip with open boundary
condition on either side. As mentioned above, the ground-state of the Hamiltonian is the
identity, of weight $h_{\rm min} = 0$, leading to an effective central charge
$c_{\rm eff} = -2$. The general formula (\ref{Astrip}) for the strip gives the
correct result (\ref{podd}).

This is a very unusual situation. Although the dimer model is originally defined on a
cylinder, it shows the finite-size corrections expected on a strip, and must really
be viewed as a model on a strip.

For $N$ even, the problem of having two arrows pointing from and
to neighbor sites does not arise, however the arrows one obtains
do not define spanning trees, but rather a combination of loops
wrapped around the cylinder and tree branches attached to the
loops. Each loop has two possible orientations. The one-to-one
correspondence between the oriented loops combined with tree
branches from one side and dimer configurations from the other
side can be established as above. The enumeration of the loop-tree
configurations needs a generalization of the Kirchhoff theorem, $Z
= \det\tilde{\Delta}$ where $\tilde{\Delta}$ is the discrete
Laplacian with antiperiodic boundary conditions. In the continuous
limit, this leads to the free theory of antiperiodic Grassmann
fields which, in turn, gives $h_{\rm min}=-1/8$ \cite{Saleur}.
{}From the general formula (\ref{Acyl}) for the cylinder, we see
that the finite-size correction (\ref{peven}) is again consistent
with $c=-2$.

The above interpretation of the peculiarities of the dimer model shed a new light on
old calculations by Ferdinand of the partition function of the dimer model on a $M
\times N$ torus \cite{Ferdinand}. When $M$ and $N$ are both even, the universal
part of the partition function equals
\be
Z_{even,even} = {\theta_2^2 + \theta_3^2 + \theta_4^2 \over 2\eta^2}(q),
\ee
in the limit $M,N \to \infty$ with fixed ratio $q=\exp{(-2\pi{M \over N})}$. This
is exactly equal to the partition function $Z_{c=-2}(q) = |\chi_{-1/8}|^2 +
2|\chi_0 + \chi_1|^2 + |\chi_{3/8}|^2 $ of the $c=-2$ rational conformal theory
developed in \cite{Gaberdiel}, confirming the value of $c=-2$ (and the value
$h_{\rm min}=-1/8$ found for the even cylinder). In case one of $M$ or $N$ is
odd, the partition functions are
\be
Z_{odd,even} = {\theta_2 \over 2\eta}(\sqrt{q}), \quad Z_{even,odd} = {\theta_4 \over
2\eta}(\sqrt{q}).
\ee
These are cylinder partition functions of the same $c=-2$ conformal theory \cite{ruelle},
in agreement with the view that a periodic dimension of odd size is actually not
periodic, and correspondingly tori with one odd dimension are in fact cylinders.

We thank F.Y. Wu for valuable comments. V.P. and P.R. would like
to thank the hospitality of the Laboratory of Statistical and
Computational Physics in Academia Sinica (Taiwan), where the most
significant part of this work was done. This work was supported by
National Science Council (Taiwan) under Grant Nos. NSC
93-2112-M001-027 \& NSC 94-2119-M-002-001, Academia Sinica
(Taiwan) under Grant No. AS-91-TP-A02, and by the Belgian Fonds
National de la Recherche Scientifique.


\end{document}